\documentclass[reprint,amsmath,amssymb,aps,pre,eqsecnum,longbibliography]{revtex4-1}

\usepackage{graphicx}
\usepackage{subfigure}
\usepackage{bm}
\usepackage{amssymb}
\usepackage{amsmath}

\usepackage{ulem}

\usepackage[utf8]{inputenc}   

\usepackage{hyperref}

\usepackage{color,soul}  
\setstcolor{red}         

\begin{document}
\title{
Continuous time random walks and Fokker-Planck equation in expanding media}
\author{F. Le Vot and S. B. Yuste}
\affiliation{
 Departamento de F\'{\i}sica and Instituto de Computaci\'on Cient\'{\i}fica Avanzada (ICCAEx    ) \\
 Universidad de Extremadura, E-06071 Badajoz, Spain
 }

\begin{abstract}

We consider a continuous random walk  model for describing normal as well as anomalous diffusion of particles subjected to an external force when these particles diffuse in a uniformly expanding (or contracting) medium.
A general equation that relates the probability distribution function (pdf) of finding a particle at a given position and time to the single-step jump length and waiting time pdfs is provided. The equation takes the form of a  generalized Fokker-Planck equation when the jump length pdf of the particle has a finite variance. This generalized equation becomes a fractional Fokker-Planck equation in the case of a heavy-tailed waiting time pdf.
These equations allow us to study the relationship between expansion, diffusion and external force.
We establish the conditions under which the dominant contribution to transport stems from the diffusive transport rather than from the drift due to the medium expansion.
We find that anomalous diffusion processes under a constant external force in an expanding medium described by means of our continuous random walk  model are not Galilei invariant,  violate the generalized Einstein  relation, and lead to propagators that are qualitatively different from the ones found in a static medium.
Our results are supported by numerical simulations.

\end{abstract}


\maketitle


\section{Introduction}

Diffusion phenomena under the influence of an external force is a main topic in the field of applied stochastic processes. Shortly after Einstein published its celebrated 1905's paper on Brownian motion, Smoluchowski developed his own approach in which effects of external forces were included. Later on, in the mid-1930s,  Ornstein and Uhlenbeck  improved the Langevin's approach to Brownian motion and considered explicitly the case of a Brownian particle subjected to a harmonic potential. A few years later, Kramers studied the problem of a Brownian particle in a force field as a way to understand chemical reaction kinetics. Since then, the number and variety of works on stochastic processes under the influence of an external force is enormous \cite{Gardiner2010, Gillespie2012}.

In the standard Brownian motion without external forces, the mean square displacement of the Brownian particle is proportional to $t^\alpha$ with $\alpha=1$.  However, in many diffusive processes in Physics, Biology, Chemistry, Finance, \ldots  one finds either $\alpha<1$ (subdiffusive processes) or $\alpha>1$ (superdiffusive processes)
\cite{AnotransBook,Bouchaud1990,Metzler2000,Sokolov2005a,Dix2008,Metzler2014b}.  Furthermore,  many of these anomalous diffusion processes take place under the influence of external fields \cite{AnotransBook,Barkai1998,Metzler1999PRL,Henry2010,Langlands2010}.  A convenient model to study these anomalous diffusion processes is the so-called Continuous Time Random Walk (CTRW) model \cite{MontrollWeiss,Barkai2012,Kutner2017}.  In this model, both  the length $\Delta y$ of the jumps of the walkers, and the waiting time $\Delta t$  between jumps, are random variables. This is the model we use in this paper.

The vast majority of works on diffusion processes assume that the medium in which the particles diffuse is ``static'', that is, it is assumed that the distance between two static and unforced walkers does not change with time.
However, this is no longer true for expanding (or contracting) media.
There are many examples in biology, fluids, chaotic systems, and cosmology where stochastic transport takes place in an expanding medium.
It turns out that the expansion of the medium has a strong influence on diffusive transport and on encounter-controlled particle reactions. \cite{Murray2003,Crampin1999,Crampin2001,Berezinsky2006,Yates2014,Averbukh2014,Simpson2015b,Yuste2016,LeVot2017}

An example of an expanding medium is the universe. It turn out that, in some cases, this expansion can be relevant for the correct description of some cosmological diffusion processes.
A nice example of this is the  diffusion of high energy cosmic rays due to extragalactic inhomogeneous magnetic fields \cite{Berezinsky2006,Berezinsky2007,Kotera2008,Batista2014}.
On the other hand, in Biology, it is well know that the growth of tissues due to cell division can be very fast, in particular at the embryonic stages. For example, the gut of some vertebrates during the first days of life, or the size of the \textit{alligator  mississippiensis} embryo,  grow exponentially \cite{Binder2008,Murray2003}.
Such fast growth processes have important consequences (see Ref.~\cite{Yates2014} and references therein).
For example, according to the French flag model, the gradient profiles of the diffusive morphogens, and the resulting spatial patterning during embryogenesis, is largely modified by the growth of the embryonic tissue \cite{Fried2015}.
The interplay between diffusion, reactions  and tissue growth in the formation of biological patterns is an important topic in developmental biology \cite{Kulesa1996,Crampin1999,Murray2003,Landman2003,Kicheva2015}.
On the other hand, it is worth noting that diffusive particles in biological media are usually subjected to a large variety of interactions, which usually gives rise to anomalous diffusion \cite{Dix2008,Metzler2014b,Hofling2013}.
In many cases  the CTRW model provides a reasonable description of such anomalous diffusion processes \cite{Metzler2014b,Hofling2013}.
For example,  recently Tan et al. \cite{Tan2018} have used a variant of the CTRW model to describe the anomalous diffusion dynamics of surface water around proteins.

In many cases the growth of the medium is known to be uniform: the distance   between two points of the medium, separated initially by a given distance, increases (or decreases) with time in a way that is independent of the position on which they are placed.
There are many examples of this kind of expansion in biological systems; see Refs.~\cite{Murray2003,Binder2008,Gerlee2013}.
On the other hand, the expansion of the universe is uniform on large scales,  the growth being exponential for a dark-energy dominated universe, and power-law in both matter and radiation dominated universes \cite{Ryden2003}.

The most usual approach for studying diffusion processes on growing media is the macroscopic, continuum description based on the use of partial differential equations for modeling the space-time evolution of the density of diffusing agents \cite{Murray2003,Simpson2015b}.
A more recent alternative approach is based on  microscopic/mesoscopic descriptions in which the starting point is the stochastic  movement of the individual agents, often modeled as random walkers \cite{Baker2010,Yates2014,Ross2016,Yuste2016,Angstmann2017}.
In particular,  the Continuous-Time Random-Walk (CTRW) model has been used in Ref.~\cite{LeVot2017} to  derive a diffusion equation for anomalous diffusive processes in uniformly expanding media.
The main aim of this paper is to generalize this study to the case in which the diffusive particles are subjected to an external force as well as to find the corresponding Fokker-Planck equation (FPE).

The plan of the paper is the following. In Sec.~\ref{Sec1} we introduce a general formalism to describe CTRW processes under an external force in uniformly expanding media.
In Sec.~\ref{Sec2} we focus on the specific cases of Brownian and subdiffusive random walks and we find the corresponding FPEs when the walkers are in an external force field. We use this equation to discuss the effects of a constant force field on the diffusion properties of particles in media with power-law or exponential growth.   We conclude with a brief summary in Sec.~\ref{Sec_Conclusions}.

\section{External force and continuous time random walks in a uniformly expanding medium}
\label{Sec1}

The CTRW model \cite{MontrollWeiss} with bias is a standard tool for describing diffusion processes in the presence of an external force. In this model, the walkers move by means of jumps. The  displacement $y-y'$ due to the jumps (instantaneous jumps from $y'$ to $y$)  and  the time elapsed between jumps $t-t'$  is drawn from a probability density function (pdf) $\psi^{*}(y,y',t,t')$. If the displacement  $\Delta y=y-y'$ and the waiting time $t-t'$ are independent random variables, one may write $\psi^*(y,y',t,t') = \Lambda^*(y,y',t) \varphi(t,t')$, where $\Lambda^*(y,y',t)$ is the pdf of jumping at time $t$ from $y'$ to $y$ and $\varphi(t,t')$ is the pdf of waiting the time $t-t'$ between two successive jumps.  The CTRW approach is a especially convenient way of dealing with some anomalous diffusion processes. Here we generalize this approach to deal with a walk biased by the action of an external force and taking place in a uniformly growing (contracting) domain. In the next section, Sec.~\ref{Sec2}, we derive the corresponding FPE. For simplicity, we focus on the one-dimensional case. The generalization for higher dimensions is straightforward and similar to the one for a static medium.

Following the procedure of Metzler et al. \cite{Metzler1998,Metzler1999}, we include the effect of the external field  in the CTRW model by means of a direction-dependent jump length distribution $\Lambda^{*}(y,y',t)$:
 \begin{align}
\Lambda^{*}(y,y',t) =& 2 \lambda^{*}(y,y') \left[ A^{*}(y',t) \Theta(y-y') \right. \nonumber\\
&+ \left. B^{*}(y',t)  \Theta(y'-y)  \right],
\label{LambdaChoice}
\end{align}
Here
\begin{equation}
\label{laylay}
\lambda^{*}(y,y')=\lambda^{*}(y',y)=\lambda^{*}(y-y')
\end{equation}
is a symmetric pdf that determines the probability that a random walker takes a jump of size $|y-y'|$, $A^{*}(y',t)$ is the probability that the walker placed at $y'$ takes an instantaneous jump to the right at time $t$, and  $B^{*}(y',t)$ is the probability that the walker placed at $y'$ takes an instantaneous jump to the left at time $t$. Obviously, $A^{*}+B^{*}=1$.   The spatial asymmetry induced by the external force implies the inequality  between $A^{*}$ and $B^{*}$.  The forceless case is simply recovered by taking $A^{*}=B^{*}=1/2$.

Let $\Delta_n y$ be the length of the $n$th jump, $t_n$ the time in which the $n$th jump is given,   $\Delta_n t=t_n-t_{n-1}$ the waiting time of the walker for taking the $n$th step, and $y_n\equiv y(t_n^+)$ the position of the walker just after the $n$th step is taken.  If we define $y(t_n^-)$ as the position of the walker just before the $n$th step is taken, one sees that
\begin{equation}
\label{Deltany}
\Delta_n y = y(t_n^+)-y(t_n^-).
\end{equation}
Note that for a static medium $y(t_{n-1}^+)=y(t_{n}^-)$ because the walker is at rest between the $(n-1)$th jump and the $n$th jump, that is, during the time interval $t_{n-1}<t<t_n$. But if the medium is not static, the equality $y(t_{n-1}^+)=y(t_{n}^-)$  is no longer true because the particles are dragged by the expansion of the medium, i.e., by the so-called Hubble flux \cite{Ryden2003}. For this reason $y_{n+1}-y_n\neq\Delta_n y $ and $y_m\neq \sum_{n=1}^m \Delta_n y$. This implies that the usual formulation of the CTRW model, and their corresponding results, are not valid for expanding media.

The difficulties introduced in the CTRW model by the expansion of the medium can be reduced by using comoving coordinates for describing the movement of the particles.
Let $x=y(0)$ be the coordinate of a fixed point at the initial time $t=0$. Due solely to the expansion of the medium, this fixed point changes its position from $y(0)$ to $y(t)$ at time $t$. The specific relation $y=f(x,t)$ between the physical position of the point, $y$, and its comoving coordinate, $x$, depends on the kind of expansion.
If the expansion of the medium is uniform, the  physical and comoving coordinates are related by
\begin{equation}
y = a(t)\, x,
\label{y=ax}
\end{equation}
with $a(0)=1$. In the cosmological context, $a(t)$ is called the scale factor.
Note that, by construction, the comoving distance between two walkers does not change as long as neither of them jumps. This allows us to study the  CTRW model in expanding media with the tools of the standard CTRW approach in static media.

The jump pdf $\psi^{*}(y,y',t,t')$ in the physical space corresponds to a jump pdf in comoving coordinates:
\begin{equation}
\label{psixLv}
\psi(x,x',t,t')= \Lambda(x,x',t) \varphi(t-t').
\end{equation}
The two pdfs are related by the following probability conservation relation:
\begin{equation}
\label{x}
\psi^{*}(y,y',t,t') dy dt=\psi(x,x',t,t')dx dt.
\end{equation}
The function $\Lambda(x,x',t)=\Lambda^*(y,y',t) dy/dx$ is the probability density of taking a jump from $x'$ to $x$ just at time $t$.   For uniformly expanding media one has
$\Lambda(x,x',t) = a(t) \Lambda^{*}(a(t) x,a(t) x',t)$ and therefore, from Eq.~\eqref{LambdaChoice},
\begin{equation}
\begin{split}
 \Lambda(x,x',t) & = 2 \lambda(x,x',t) \left[ A \Theta(x-x') + B  \Theta(x'-x)  \right] \\ & = 2 \lambda(x,x',t) \left[ \Theta(x-x')  \left( A - B \right) + B\right] ,
\end{split}
\label{Lambda-x-t-Drift}
\end{equation}
 where $A\equiv A(x',t)=A^{*}(a(t) x',t)$,  $B\equiv B(x',t)=B^{*}(a(t) x',t)$, and
 \begin{equation}
 \label{laxlay}
\lambda(x,x',t) = a(t) \lambda^{*}(a(t) x, a(t) x').
\end{equation}
From the definition of $\lambda^*(y)$ given in Eq.~\eqref{laylay}, one sees that
\begin{equation}
\label{x}
\lambda(x,x',t)=\lambda(x',x,t)= \lambda(x-x',t).
\end{equation}

Let us define $\eta(x,t)$ as the pdf of arriving at the comoving position $x$ at time $t$. This comoving arrival density is equal to the sum of the probabilities of arriving at any other site $x'$ at $t'<t$, and then taking a jump from $x'$ to $x$ at time $t$. This function satisfies \cite{Klafter1987,Metzler2000}
\begin{equation}
\label{etaxt2}
\eta(x,t) =  \int_{-\infty}^{\infty} dx' \int_0^t dt' \eta(x',t') \psi(x,x',t,t') + \delta(x) \delta (t),
\end{equation}
where $ \delta(x) \delta (t) $  accounts for the initial condition.

Introducing Eqs.~\eqref{psixLv} and \eqref{Lambda-x-t-Drift} into Eq.~\eqref{etaxt2} and   taking the Fourier transform on both sides of the resulting equation, one finds
\begin{align}
\widehat{\eta}(k,t) =
2  &\left\{ \mathcal{F} \left[ \lambda \Theta \right] \mathcal{F} \left[ (A-B) \mathcal{L}^{-1} \left(\tilde{\eta} \tilde{\varphi} \right) \right] \right. \nonumber\\
+ &\left. \widehat{\lambda} \mathcal{F} \left[B \mathcal{L}^{-1} \left(\tilde{\eta} \tilde{\varphi} \right) \right] \right\}
+ \delta(t)
\label{Eta-k-t-Drift}
\end{align}
where $\mathcal{F}$ is the Fourier transform operator,
\begin{equation}
\mathcal{F}[f(x)]=\hat{f}(k)=\int_{-\infty}^\infty e^{-ikx} f(x) dx,
\end{equation}
 and  $\mathcal{L}$  is the Laplace transform operator
\begin{equation}
\mathcal{L}[f(t)]=\tilde{f}(s)=\int_0^\infty e^{-st} f(t) dt.
\end{equation}

The pdf $\eta(x,t)$ is closely related to the pdf $W(x,t)$ of finding a walker at position $x$ at time $t$. When $W(x,0)=\delta(x)$, the function  $W(x,t)$ is called the propagator or Green function.
Of course, the pdf $W^*(y,t)$ of finding a walker at position $y$ at time $t$ and $W(x,t)$ are related by $W^*(y,t)=W(y/a(t),t)/a(t)$.
The relationship between $\eta(x,t)$ and $W(x,t)$ is \cite{Metzler2000,Klafter2011}
\begin{equation}
W(x,t) = \int_0^t dt' \eta (x,t') \Phi(t-t')
\end{equation}
where $\Phi(t) = 1-\int_0^t dt' \varphi(t')$
is the probability that the walker does not jump during the the time interval $(0,t)$. In the Laplace space one has
\begin{equation}
\widetilde{W}(x,s) = \tilde{\eta} (x,s) \widetilde{\Phi}(s)
\label{W-Eta-x-s}
\end{equation}
with
\begin{equation}
\widetilde\Phi(s) = \frac{1-\tilde\varphi(s)}{s}.
\label{Phis}
\end{equation}
Taking the Laplace transform on both sides of Eq.~\eqref{Eta-k-t-Drift}  and making use of Eq.~\eqref{W-Eta-x-s}, we finally obtain the equation that relates $W$ to the single-step pdfs $\lambda$ and $\varphi$:
\begin{align}
\frac{\widehat{\widetilde{W}}}{\widetilde{\Phi}} -1 =
 &2 \mathcal{L} \left\{  \mathcal{F} \left[ \lambda \Theta \right] \mathcal{F} \left[ (A-B) \mathcal{L}^{-1} \left( \tilde{\varphi} \frac{ \widetilde{W}}{\widetilde{\Phi}}  \right) \right]\right. \nonumber  \\
 & + \left.\widehat{\lambda} \mathcal{F} \left[B \mathcal{L}^{-1} \left(\tilde{\varphi} \frac{\widetilde{W}}{\widetilde{\Phi}} \right) \right] \right\} .
\label{W-k-s-Drift}
\end{align}
If the medium is static and $A^*$ and $B^*$ do not depend on time, one can show that this equation is equivalent to a generalized master equation \cite{Metzler1999}.  Furthermore, if the  jump pdf $\Lambda$ depends only on the difference $x-x'$ (which implies that $A$ and $B$ are constant),  one easily recovers  the  Montroll-Weiss equation  \cite{MontrollWeiss}
\begin{equation}
\label{x}
\widehat{\widetilde{W}}(k,s)= \frac{\widetilde\Phi(s) }{1-\tilde\varphi(s) \widehat{\Lambda}(k)}
\end{equation}
from Eq.~\eqref{W-k-s-Drift}.

\section{FPE for walkers with finite jump-length variance in a uniformly expanding medium }
\label{Sec2}

Equation~\eqref{W-k-s-Drift} is valid for any jump length pdf $\lambda^*$ and any waiting time pdf $\varphi$ and for any uniformly expanding medium.  In this paper we are going to consider only cases where $\lambda^*(y)$ has a finite second moment.  L\'evy flights where $\lambda^*(y)$  has a diverging variance will be considered elsewhere.

\subsection{FPE for jump lengths with finite variance}

The jump length pdf $\lambda^{*}(y)$ we consider in this paper is symmetric and has a finite second moment that we denote by $2\sigma^2$.
We will refer to $\sigma^2$ as the semivariance. In this case  $\widehat\lambda^*(k)\sim 1-\sigma^2 k^2$ for small $k$. From Eq.~\eqref{laxlay} one finds that if $m_j^*$ is the $j$-th moment of $\lambda^*(y)$, then the $j$-th moment of $\lambda(x,t)$ is $m_j(t)=m_j^*/a^j(t)$.  In particular $m_2(t)=2\sigma^2/a^2(t)$ and, therefore,
\begin{equation}
\widehat{\lambda} (k,t) \sim 1 - k^2 \frac{\sigma^2}{a^2(t)}
\end{equation}
for small $k$.

Let $M^{*}_j$ be the $j$-th semimoment of $\lambda^{*}(y)$:
\begin{equation}
M^{*}_j = \int_{-\infty}^{\infty} y^j \lambda^{*}(y) \Theta(y) dy = \int_0^{\infty} y^j \lambda^{*}(y) dy,
\end{equation}
and let $M_j (t)$ be the $j$-th semimoment of $\lambda(x,t)$.  It is clear that $M_j (t) = M^{*}_j / a^j (t)$.
On the other hand, it's not difficult to see that
\begin{equation}
\mathcal{F}({\lambda \Theta})  \sim M_0 - i k M_1(t) - k^2 \frac{M_2 (t)}{2}.
\end{equation}
Taking into account that $\lambda^*(y)$ is an even function, one finds that $M_j(t)=m_j(t)/2$ for $j$ even. Therefore
\begin{equation}
\label{FLaThe}
\mathcal{F}({\lambda \Theta})  \sim  \frac{1}{2} - i k \frac{M_1^*}{a(t)} - k^2 \frac{\sigma^2}{2 a^2(t)}
\end{equation}
In what follows we write $M_1^*=\varepsilon \sigma$,$\varepsilon$  being  a non dimensional constant that depends on $\lambda^*$. For example, $\varepsilon=1/\sqrt{\pi}$ for the Gaussian jump distribution
\begin{equation}
\lambda^{*}(y) = \frac{1}{\sqrt{4 \pi \sigma^2}} \exp \left( - \frac{y^2}{4 \sigma^2} \right).
\label{Gaussianpdf}
\end{equation}

Substituting Eq.~\eqref{FLaThe} into Eq.~\eqref{W-k-s-Drift}, one finds
\begin{align}
s\widehat{\widetilde{W}}(k,s) - 1 = & -\mathcal{L} \left[ \frac{k^2 \sigma^2 }{a^2(t)} \mathcal{L}^{-1} \left(\tilde{\varphi} \frac{ \widehat{\widetilde{W}}}{\widetilde\Phi} \right) \right] \nonumber \\
 & - \mathcal{L} \left\{ \frac{2 i k \varepsilon \sigma  }{a(t)} \mathcal{F} \left[ (A-B) \mathcal{L}^{-1} \left( \tilde{\varphi}  \frac{ \widehat{\widetilde{W}}}{\widetilde\Phi} \right) \right] \right\}
\label{DriftDiffusionEquationFL}
\end{align}
which is equivalent to
\begin{align}
\frac{\partial W(x,t)}{\partial t} =& \frac{\sigma^2}{a^2(t)} \mathcal{L}^{-1}  \left( \frac{  \tilde{\varphi}}{\widetilde\Phi} \frac{\partial^2 \widetilde{W}}{\partial x^2} \right) \nonumber \\
&+ \frac{2 \varepsilon \sigma}{a(t)} \frac{\partial}{\partial x} \left[ (B-A) \mathcal{L}^{-1} \left(\tilde{\varphi} \frac{  \widetilde{W}}{\widetilde\Phi} \right) \right],
\label{DriftDiffusionEquation}
\end{align}
which has the form of a (generalized) Fokker-Planck equation.

\subsection{Normal FPE in expanding media}
Equation~\eqref{DriftDiffusionEquation} reduces to the normal diffusion-advection equation in an expanding medium \cite{Yuste2016} if $\varphi(t)$ is a continuous function with finite first moment, $\langle \varphi \rangle =\tau$. An example is the exponential pdf
 \begin{equation}
\varphi(t) = \exp (-t/\tau) / \tau.
\label{Exponentialpdf}
\end{equation}
In these cases $\tilde{\varphi}(s) \sim 1-\tau s+\cdots$ for small $s$. Taking into account  Eq.~\eqref{Phis}, one has $\tilde{\varphi}/\widetilde\Phi\sim 1/\tau$. Then,   Eq.~\eqref{DriftDiffusionEquation} becomes
\begin{equation}
\frac{\partial W(x,t)}{\partial t} = \frac{\mathfrak{D}}{a^2(t)} \frac{\partial^2 W}{\partial x^2} - \frac{1}{a(t)} \frac{\partial}{\partial x} \left[ v(x,t) W(x,t) \right],
\label{EDNDME}
\end{equation}
where $\mathfrak{D}\equiv\mathfrak{D}_1=\sigma^2/\tau$ is the diffusion coefficient and $v(x,t)$ is given by
\begin{align}
\label{vxt}
v(x,t)\equiv &\frac{2\varepsilon \sigma}{\tau} \left[ A(x,t)-B(x,t)\right]\\
=& \frac{2\varepsilon \sigma}{\tau} \left[ A^{*}(y,t)-B^{*}(y,t) \right] \equiv v^{*}(y,t).
\label{vy}
\end{align}
From Eq.~\eqref{LambdaChoice} one finds that the mean value of the displacement $z$  after a single jump  is
\begin{align}
\langle z\rangle&= \int_{-\infty}^\infty z \Lambda(z+y,y,t) dz \nonumber\\
&=2[A^{*}(y,t)-B^{*}(y,t)] \int_0^\infty z\lambda^*(z) dz \nonumber\\
&= 2[A^{*}(y,t)-B^{*}(y,t)] \varepsilon \sigma .
\end{align}
Comparing this equation with Eq.~\eqref{vxt}, we see that $v(x,t)$ is just the mean displacement of the walker after a single jump, $\langle z\rangle$, divided by the mean time $\tau$ employed by the walker for taking a jump. Then  $v(x,t)$ can be interpreted as the net drift velocity of the walkers due to the asymmetry of
the jump distribution $\Lambda$.  These results were obtained in Ref.~\cite{Yuste2016} but by means of a  Chapman-Kolmogorov approach.

\subsection{Fractional FPE in expanding media}

The waiting time pdfs of subdiffusive CTRWs  are  heavy-tailed distributions:  $\varphi(t) \sim   t^{-1-\alpha}$ with $0<\alpha<1$ for long times \cite{Metzler2000}. In particular, for
\begin{equation}
\label{phiAsinPare}
\varphi(t)\sim \frac{\alpha}{\Gamma(1-\alpha)} \frac{\tau^\alpha}{t^{1+\alpha}}
\end{equation}
one has    $\tilde{\varphi}(s) \sim 1 -\tau^{\alpha} s^{\alpha}$ when $s \to 0$.  In this case the mean value of $\varphi(t)$ does not exist and $\tau$ merely represents a typical time related to the decay-rate of $\varphi(t)$.
From Eq.~\eqref{Phis} one finds $\tilde\varphi/\widetilde\Phi \sim s^{1-\alpha}/\tau^\alpha$ for small $s$. Inserting this expression into  Eq.~\eqref{DriftDiffusionEquation} one obtains
\begin{align}
\frac{\partial W(x,t)}{\partial t} =& \frac{\sigma^2}{a^2(t)\tau^\alpha} \frac{\partial^2}{\partial x^2} \left[ \mathcal{L}^{-1}  \left( s^{1-\alpha} \tilde{W} \right) \right] \nonumber \\
&+ \frac{2 \varepsilon \sigma}{a(t)\tau^\alpha} \frac{\partial}{\partial x} \left[ (B-A) \mathcal{L}^{-1} \left( s^{1-\alpha} \tilde{W} \right) \right] .
\label{FFPE28}
\end{align}
But  $\mathcal{L}^{-1} [ s^{1-\alpha} \tilde{f}(s) ] = {_0}\mathcal{D}^{1-\alpha}_t f(t)$,
where the operator ${_0}\mathcal{D}^{1-\alpha}_t$ is the Gr\"unwald-Letnikov fractional derivative of order $1-\alpha$ \cite{Podlubny1999}. This operator is equivalent to the Riemann-Liouville fractional derivative
\begin{equation}
  {_{~~0}^{RL}}D^{1-\alpha}_t f(t) \equiv \frac{1}{\Gamma(\alpha)} \frac{\partial}{\partial t} \int_0^t du \frac{f(u)}{(t-u)^{1-\alpha}}
\end{equation}
if $f(u)$ is continuous and $df/du$ is integrable in the interval $[0,t]$ with $0<u<t$ \cite{Podlubny1999}.

In terms of the Gr\"unwald-Letnikov derivative the FPE \eqref{FFPE28}  becomes the fractional FPE
\begin{align}
\frac{\partial W(x,t)}{\partial t} =& \frac{\mathfrak{D}_{\alpha}}{a^2(t)} \frac{\partial^2 }{\partial x^2} ~ {_0}\mathcal{D}^{1-\alpha}_t W(x,t) \nonumber \\
 &- \frac{1}{a(t)} \frac{\partial}{\partial x} \left[ v_{\alpha}(x,t) ~ {_0}\mathcal{D}^{1-\alpha}_t W(x,t) \right] ,
\label{EDADME}
\end{align}
where $\mathfrak{D}_{\alpha}=\sigma^2 / \tau^{\alpha}$ is the anomalous diffusion constant and
\begin{align}
\label{vxt_alpha}
v_{\alpha}(x,t)
\equiv & \frac{2 \varepsilon \sigma}{\tau^{\alpha}} [A(x,t)-B(x,t)] \\
=&
\frac{2 \varepsilon \sigma}{\tau^{\alpha}} [A^*(y,t)-B^*(y,t)]
\equiv v^*_\alpha(y,t) .
\label{vy_alpha}
\end{align}
This definition of  $v_{\alpha}$ is just a generalization of the definition of $v$ of Eq.~\eqref{vxt} for any anomalous diffusion exponent $\alpha \in (0,1]$.
However, $v_{\alpha}$ is not a drift velocity (it is not even a velocity); it is just a measure of the walker's preference to move to  a given  direction.

\subsection{Force and bias}
The existence of an external force leads to $A^*\neq B^*$. For example, $A^*>B^*$ if the force pushes the particle to the right. For a static medium, the relationship  between the external force $F^*(y,t)$ and the asymmetry of the jumps of the walker [asymmetry accounted for by the quantity $v_\alpha^*\propto A^*-B^*$ in Eq.~\eqref{EDADME}] is well known for normal diffusion as well as for subdiffusive processes described by the CTRW model  \cite{Metzler1999PRL}, namely,
\begin{equation}
\label{vFxi}
v_\alpha^*=\frac{F^*}{\xi_\alpha}
\end{equation}
where $\xi_\alpha$ is a generalized friction factor (or generalized drag coefficient).  Taking into account the generalized Stokes-Einstein-Smoluchowski relation  \cite{Metzler1999PRL}
\begin{equation}
\label{SESrelation}
\mathfrak{D}_{\alpha} \xi_\alpha=k_B T
\end{equation}
and Eqs.~\eqref{vy_alpha} and \eqref{vFxi}, one finds
\begin{equation}
\label{FsAB}
\frac{F^*\sigma}{k_BT}=2\varepsilon (A^*-B^*).
\end{equation}
This equation, or equivalently
\begin{equation}
\label{FsABbis}
A^*-B^*= \frac{F^*\sigma}{2\varepsilon\mathfrak{D}_{\alpha} \xi_\alpha},
\end{equation}
relates the asymmetry $A^*-B^*\neq 0$ of the jump probabilities to the external force.

Note that $|A^*-B^*|$ cannot be larger than one, which implies that all the forces larger than, or equal to $2\varepsilon \mathfrak{D}_{\alpha} \xi_\alpha/\sigma$ have the same effect on the random walker, i.e., the effect of the forces saturates at $|F|=2\varepsilon \mathfrak{D}_{\alpha} \xi_\alpha/\sigma=2\varepsilon \sigma \xi_\alpha/\tau^\alpha$.

It is sensible to expect that the effect of the force on the bias of the walker's jump probability is independent of the kind of medium, static or expansive, in which the walker moves.  In other words, one expects Eq.~\eqref{FsAB}, or equivalently Eq.~\eqref{vFxi}, $v_{\alpha}(x,t)=v_{\alpha}^*(y,t)=F^*(y,t)/\xi_\alpha$, to hold true for expanding media.  This in turns implies that the fractional FPE~\eqref{EDADME} for walkers subjected to an external force  in a uniformly expanding medium can be written as
\begin{align}
\frac{\partial W(x,t)}{\partial t} =& \frac{\mathfrak{D}_{\alpha}}{a^2(t)} {_0}\mathcal{D}^{1-\alpha}_t \left[ \frac{\partial^2 W}{\partial x^2} \right] \nonumber \\
&- \frac{1}{a(t)} \frac{1}{\xi_{\alpha}} \frac{\partial}{\partial x} \left[ F(x,t) ~ {_0}\mathcal{D}^{1-\alpha}_t W(x,t) \right] ,
\label{EDADME_Force}
\end{align}
where $F(x,t)=F^{*}(y=a(t)x,t)$.  For $a(t)=1$ (static medium), this equation is just the one obtained by Henry et al. in Ref.~\cite{Henry2010}.

\subsection{Simulation of continuous  time random walkers in an expanding medium}

The computer simulation of continuous  time random walkers in expanding media requires specifying how the walkers jump and how the expansion of the medium modifies the position of the walkers.

The simulation of the jumps is carried out as for a static medium. The walker  jumps at times $t_m$ with jumps $\Delta_m y=y(t_m^+)-y(t_m^-)$. These quantities are random variables: the time interval between jumps, $\Delta_m t=t_m-t_{m-1}$, is drawn from a waiting time pdf $\varphi(\Delta t)$ and the jump $\Delta_m y$ is drawn from the jump length pdf $\bar{\Lambda}^*(\Delta y=y-y',y',t)\equiv \Lambda^{*}(y,y',t)$, that is, from [see Eq.~\eqref{LambdaChoice}]
 \begin{equation}
\bar{\Lambda}^*= 2 \lambda^{*}(\Delta y) \left[ A^{*}(y',t) \Theta(\Delta y) + B^{*}(y',t)  \Theta(-\Delta y)  \right].
\label{LambdaPDFbis}
\end{equation}
In our simulations we  use the  Gaussian jump-length pdf of Eq.~\eqref{Gaussianpdf} with $\sigma^2=1/2$.  When simulating normal diffusive particles, we use the waiting time pdf of Eq.~\eqref{Exponentialpdf} with $\tau=1$, whereas we use the Pareto pdf
\begin{equation}
\label{varphiSimu}
\varphi(t)=\frac{\alpha/t'}{(1+t/t')^{1+\alpha}},
\end{equation}
with $\alpha=1/2$ and $t'=1/\pi$, in our simulations of anomalous diffusive particles.
Comparing Eq.~\eqref{varphiSimu} with Eq.~\eqref{phiAsinPare} one sees  that $\tau=1$ in this case.   Note that we always simulate random walkers with $\sigma^2=1/2$ and $\tau=1$, which implies $\mathfrak{D}_\alpha=\sigma^2/\tau^\alpha=1/2$.

The expansion of the medium introduces some difficulties in the simulation of the CTRW that we must handle carefully. As we discussed at the beginning of Sec.~\ref{Sec1}, $y(t_{n-1}^+)\neq y(t_{n}^-)$ because the medium expands between jumps. Therefore,  $y(t_n^+)- y(t_{n-1}^+)\neq \Delta_n y$  [recall that  $\Delta_n y=y(t_n^+)-y(t_n^-)$; see Eq.~\eqref{Deltany}].
On the other hand, because there is no jump between $t_{n-1}^+$  and $t_{n}^-$, the comoving position $x$ of the particle does not change during this time interval, that is, $x(t_{n-1}^+)=x(t_{n}^-)$. This implies $y(t_{n-1}^+)/a(t_{n-1})=y(t_n^-)/a(t_n)$. From this equation and Eq.~\eqref{Deltany} one finds that the position of the walker just after the $n$th jump is given by
\begin{equation}
   \label{x}
y(t_{n}^+)=\frac{a(t_n)}{a(t_{n-1})} y(t_{n-1}^+)+\Delta_n y.
\end{equation}
The position of the walker for any time $t$ with $t_n<t<t_{n+1}$  is simply given by $y(t)=a(t) y(t_{n}^+)/a(t_n)$.

\section{Diffusion under a constant force in expanding media}

In this section, we consider the case of diffusive particles subjected to a constant external force in a uniformly expanding medium. We will see that the FPEs introduced in Sec.~\ref{Sec2} describe accurately this problem and, along the way, we will discover some interesting results on the relationship between the external force, the expansion of the medium, and the waiting time pdf of the particles.

\subsection{Normal diffusion under a constant force}
\label{Sec:ConstantDrift}

For normal diffusion, $\alpha=1$,  and constant force, $F=\xi v$,  Eq.~\eqref{EDNDME} becomes
\begin{equation}
\frac{\partial W (x,t)}{\partial t} = \frac{\mathfrak{D}}{a^2(t)} \frac{\partial^2 W(x,t)}{\partial x^2} - \frac{v}{a(t)} \frac{\partial W(x,t)}{\partial x},
\label{BrownianDriftDiffusionEquation}
\end{equation}
which was recently obtained by Yuste et al.~\cite{Yuste2016} by using an approach based on a generalized Chapman-Kolmogorov equation. For the sake of completeness, we provide here some results for this case.

Applying the Fourier transform operator  to Eq.~\eqref{BrownianDriftDiffusionEquation} one finds
\begin{equation}
\log   \frac{ \widehat{W} (k,t)}{\widehat{W_0} (k)}   = -\mathfrak{D} k^2 T_2(t) - i v k  T_1(t),
\end{equation}
where $\widehat{W_0} (k) = \mathcal{F} \left[W(x,0)\right]$ and
\begin{equation}
\label{x}
T_\mu(t) = \int_0^t \frac{dt'}{a^{\mu} (t')}.
\end{equation}
Therefore $\widehat{W} (k,t) =\widehat{W_0} (k) \exp [-\mathfrak{D} k^2 T_2(t) - i v k T_1 (t) ]$.
For the initial condition $W(x,0)=\delta(x)$, one has $\widehat{W_0} (k)=1$, and then one easily obtains the propagator (or Green's function):
\begin{equation}
\label{WxtNorm}
W(x,t) =
\mathcal{N}\left\{\langle x\rangle,2\sigma_x^2\right\}
\equiv\frac{1}{\sqrt{4 \pi \sigma_x^2(t)}} \exp \left[ - \frac{(x - \langle x\rangle)^2}{4\sigma_x^2(t)} \right]
\end{equation}
where
\begin{equation}
\label{xvT1}
\langle x\rangle =  v T_1
\end{equation}
is the first moment of the position of the walker and
\begin{equation}
\text{Var}(x)\equiv 2\sigma^2_x(t)\equiv \langle x^2\rangle -\langle x\rangle^2  =  2 \mathfrak{D} T_2
\label{VarxT2}
\end{equation}
is the variance.
The propagator is a Gaussian function with its characteristic symmetric ``bell curve'' shape  centered at   $\langle x\rangle$ and of width proportional to $\sigma_x(t)$.  From these results written in terms of comoving coordinates one can  straightforwardly obtain the corresponding ones in physical coordinates.  In particular,
\begin{equation}
\label{avey}
\langle y \rangle = a(t) \langle x(t) \rangle ,
\end{equation}
and
\begin{equation}
\label{vary}
\mathrm{Var} ( y ) \equiv 2\sigma^2_y(t)= 2 a^2(t) \sigma_x^2(t) .
\end{equation}

From Eqs.~\eqref{WxtNorm}--\eqref{vary} one sees that the  behavior of the propagator and its moments is determined by the behavior of the  conformal times $T_1$ and $T_2$, which in turn depends on how the medium expands.  In this paper we consider the cases of  power-law expansion and exponential expansion.

\subsubsection{Normal diffusion, constant force, and power-law expansion}

The scale factor for the power-law expansion we consider is
\begin{equation}
\label{aPowLaw}
a(t)=\left(\frac{t+t_0}{t_0}\right)^\gamma.
\end{equation}
In this case
\begin{equation}
 T_{\mu}(t) =
 \begin{cases}
   t_0 \log \left( \dfrac{t+t_0}{t_0} \right) & \text{if} \quad  \mu \gamma= 1, \\[3mm]
  \dfrac{t_0}{\mu \gamma -1} \left[ 1 - \left(\frac{t+t_0}{t_0}\right)^{1- \mu \gamma } \right] & \text{if} \quad   \mu \gamma\neq 1.
\end{cases}
\end{equation}
Note that $\lim_{t\to\infty}T_1(t)=T_1^\infty=t_0/(\gamma-1)$ when   $\gamma>1$, whereas $\lim_{t\to\infty} T_2(t)=T_2^\infty=t_0/(2\gamma-1)$ if $\gamma>1/2$. We can distinguish several regimes with qualitatively different behaviors:

\paragraph{}  For $\gamma>1$ the propagator  $W(x,t)$ goes to a stationary Gaussian function $W^\infty(x)$ when $t\to\infty$. This stationary distribution is given by  Eq.~\eqref{WxtNorm} with $\langle x\rangle =v T_1^\infty$ and $\sigma_x^2(\infty)=  \mathfrak{D} T_2^\infty$.   On the other hand, $a(t)\sim t^\gamma$ for $t\to\infty$. Therefore, from Eqs.~\eqref{avey} and \eqref{vary}, one finds $\langle y\rangle \sim t^\gamma$ and $\sigma_y^2(t) \sim t^{2\gamma}$ for long times. Note that this is how the distance and the square of the distance between two static points grow due to the expansion of the medium. Therefore, we conclude that for  power-law expansions with $\gamma>1$ (fast power-law expansions) the expansion of the medium is eventually the only relevant factor in the spreading of particles, being  negligible the contribution of their diffusive movement.  This is an expansion dominated regime.

\begin{figure}[t]
\begin{center}
\includegraphics[width=0.45\textwidth]{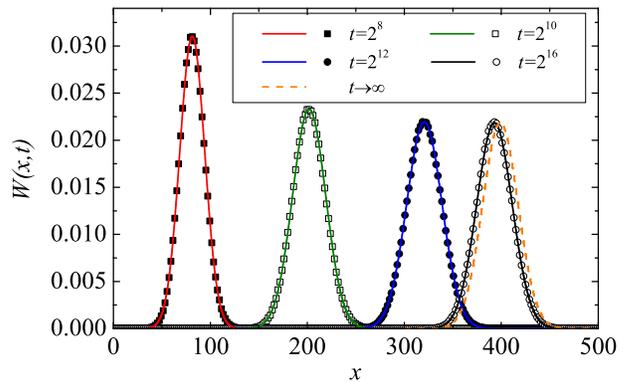}
\end{center}
\caption{\label{Fig:NormalDriftGamma2}
Propagator $W(x,t)$  for normal diffusive particles under a constant force  in a power-law expanding medium with $\gamma=2$ and $t_0=10^3$. The jump length distribution $\lambda^*(y)$ and waiting time pdf $\varphi(t)$ are given by Eq.~\eqref{Gaussianpdf} and Eq.~\eqref{Exponentialpdf}, respectively, with  $\sigma^2=1/2$ and $\tau=1$. The probability of jumping to the right due to the force is  $A=3/4$ and, therefore,  $v=1/\sqrt{2\pi}$. The symbols are simulation results for $t=2^{8}$ (filled squares)  $t=2^{10}$ (open  squares), $t=2^{12}$ (filled circles) and $t=2^{16}$ (open circles).  The solid lines are the corresponding theoretical results given by Eq.~\eqref{WxtNorm}.  The broken line is the limit stationary propagator $W^\infty(x)$.
}
\end{figure}

In Fig.~\ref{Fig:NormalDriftGamma2} we show $W(x,t)$ for a power-law expanding medium with $\gamma>2$ for four different times. For the largest time, $t=2^{16}$, the propagator is close to the final stationary propagator $W^\infty(x)$.  Note that the width of the propagators are very similar for the four times, albeit their positions are clearly different.  This is due to the fact that $T_2(t)$ converges to its final value faster than  $T_1(t)$.

\paragraph{} For $\gamma=1$ the propagator $W(x,t)$  is quasi-stationary: the average position of the walkers grows logarithmically, $\langle x\rangle \sim \log t$, whereas the variance goes to a constant value,  $\sigma_x^2(t\to\infty)\to  \mathfrak{D} t_0$.  In physical space, the behavior of $\langle y\rangle$ and $\sigma_y^2(t)$  is the same as for $\gamma>1$, save for the logarithmic factor $\log t$ in $\langle y\rangle$.

\paragraph{}  For $1/2<\gamma<1$ one finds  $\langle y\rangle \sim vt$ and $\sigma_y^2(t)\sim t^{2\gamma}$ for long times, i.e.,  the mean position is determined by the external force whereas the width of the propagator  stems from the expansion of the medium.

\paragraph{}  For $\gamma=1/2$ one obtains  $\langle y\rangle \sim vt$ and $\sigma_y^2(t)\sim t \log t$ for long times. Save for the logarithmic factor, this is the same behavior as for $\gamma<1/2$.

\paragraph{}For $\gamma<1/2$ (including contractive media where $\gamma<0$) one finds  $\langle y\rangle \sim vt$ and $\sigma_y^2(t)\sim t$ for long times.  These are just the results corresponding to a static medium.  In this regime the effect of the expansion of the medium on the spreading of the particles is negligible. This spreading is mainly determined by the external force and the diffusion process.

\subsubsection{Normal diffusion, constant force, and exponential expansion}

The scale factor of the uniform exponential expansion we  consider is
 \begin{equation}
 \label{atexp}
a(t)=\exp (Ht)\equiv \exp[t/t_H].
 \end{equation}
In the context of Cosmology, $H$ is called the Hubble parameter and
\begin{equation}
\label{tHdef}
t_H=\frac{1}{H}
\end{equation}
is the Hubble time \cite{Ryden2003}. It should be noted that the Hubble time is usually defined in Cosmology only for $H>0$.  Our definition of $t_H$ implies a  negative Hubble time when the medium is contracting.

From Eqs.~\eqref{atexp} and \eqref{atexp} one easily finds that
 \begin{equation}
 T_{\mu}(t) =  \dfrac{1-\exp \left(- \mu H t \right)}{\mu H}.
 \end{equation}
We can distinguish three different regimes:

\paragraph{} For a static medium, $H=0$, one has  $T_{\mu}(t)=t$. In this case, Eqs.~\eqref{WxtNorm}, \eqref{xvT1}, and \eqref{VarxT2}  yield the corresponding well-known Gaussian propagator
$W(x,t)=\mathcal{N}\left\{vt,2 \mathfrak{D}t\right\}$.

\paragraph{}  For $H>0$ one has $T_1^\infty= t_H$ and $T_2^\infty=t_H/2$ and the propagator $W(x,t)$ eventually reaches the stationary state $W(x,\infty)=\mathcal{N}\left\{ v t_H,\mathfrak{D} t_H\right\}$ for long times, which implies $\langle y\rangle \sim v t_H \exp(t/t_H)$ and 2$\sigma_y^2(t)\sim \mathfrak{D}t_H\exp(2t/t_H) $. This means that the diffusion process is completely dominated by the expansion of the medium.

\paragraph{} For the contractive case, $H<0$,  one finds that $T_\mu(t)$ goes as $\exp(-\mu H  t)/(-\mu H)$ for large $t$,  and therefore $\langle y\rangle \to -v t_H$ and $2\sigma_y^2(t)\to -\mathfrak{D}  t_H$ for long times. Therefore, the distribution of normal diffusive  particles  in physical coordinates eventually reaches the Gaussian stationary state $W^*(y,\infty)=\mathcal{N}\left\{-v t_H,-\mathfrak{D} t_H \right\}$.

\subsection{Anomalous diffusion under a constant force}
\label{Sec:Anomalous_Constant}

For a static medium one can obtain the solution  $W_\alpha(x,t)$  for subdiffusive particles from the corresponding solution $W_1(x,t)$ for Brownian diffusive particles via the subordination formula \cite{Barkai2001,Metzler2000}
\begin{equation}
\label{Wsubor}
W_\alpha(x,t)=\int_0^\infty r(t',t) W_1(x,t') dt' ,
\end{equation}
where
 \begin{equation}
 \label{x}
 r(z,t)=  \frac{1}{\alpha} \left(\frac{\mathfrak{D}_\alpha}{\mathfrak{D}}\right)^{1/\alpha}  \frac{t}{z^{1+1/\alpha}}\, l_\alpha\left(\frac{\mathfrak{D}_\alpha^{1/\alpha} t}{\mathfrak{D}^{1/\alpha} z^{1/\alpha}} \right)
 \end{equation}
and $l_\alpha$ is the one-sided L\'evy stable probability density whose Laplace transform is
\begin{equation}
\label{x}
\tilde l_\alpha(s)= \exp(-s^\alpha).
\end{equation}
Equation \eqref{Wsubor} can be deduced considering the subdiffusive diffusion process as a process subordinated to a Brownian random walk \cite{Sokolov2005a,Barkai2000b,Klafter2011,Metzler2000}:
\begin{equation}
\label{x}
W_\alpha(x,t)=\sum_n W_1(x,n) \chi_n(t)
\end{equation}
where  $W_1(x,n)$ is the probability density function of finding the (normal) diffusive particle at position $x$ after $n$ steps, and $\chi_n(t)$ is the probability to take exactly $n$ steps up to time $t$. Unfortunately, this approach is not valid when the medium grows because, in this case, the probability of finding the particle at a given position $x$ after $n$ steps depends also on the times at which the steps where taken.

The above discussion shows that finding exact solutions of the fractional FPE~\eqref{EDADME} for expanding media is not easy. Fortunately, Eq.~\eqref{EDADME} can be solved numerically. Besides, useful information about the expansion-diffusion process can be extracted from the first moments of $W(x,t)$, which can be directly obtained from Eq.~\eqref{EDADME}.

\subsubsection{Numerical solution of the fractional FPEs}

In what follows, we solve the fractional FPE for expanding media, Eq.~\eqref{EDADME}, by means of the fractional Crank-Nicolson method developed in Ref.~\cite{Yuste2006}. This is  a convergent and unconditionally stable  finite difference method in which the space and time are discretized in intervals of size $\Delta x$ and $\Delta t$, respectively. Its accuracy is of order $(\Delta x)^2$ and $\Delta t$.

\begin{figure}[t]
{\includegraphics[width=0.48\textwidth]{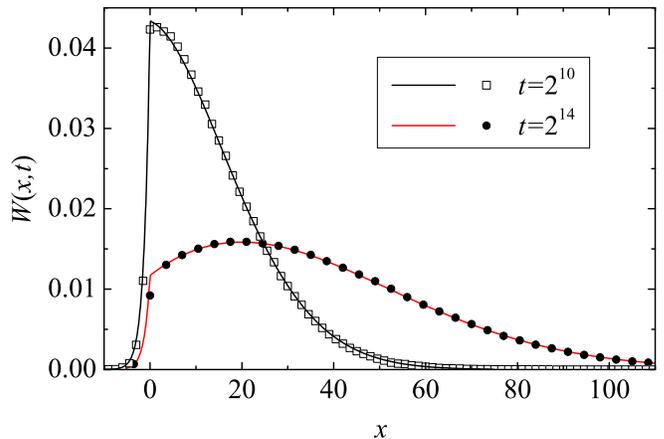}}\hfill
\caption{Propagator $W(x,t)$  for subdiffusive random walkers ($\alpha=1/2$ and $\mathfrak{D}_{\alpha}=1/2$)  in an exponentially expanding medium ($t_H=10^{4}$) subjected to an external force field ($A-B=1/2$, $v_\alpha= 1/\sqrt{2\pi}$) at times   $t=2^{10}$ (squares)   and  $t=2^{14}$ (circles). The lines represent the numerical solution of Eq.~\eqref{EDADME}  obtained by means of the fractional Crank-Nicolson method with $\Delta x=0.1$  and $\Delta t=0.1$ for $t=2^{10}$,  and with $\Delta x=0.1$  and $\Delta t=1$ for $t=2^{14}$.
The symbols are  simulation results for $10^6$ realizations where
the jump length pdf of the walkers is the same as in Fig.~\ref{Fig:NormalDriftGamma2}, and  their waiting time pdf is the Pareto distribution \eqref{varphiSimu}.
}
\label{Fig:DriftWxt}
\end{figure}

In Fig.~\ref{Fig:DriftWxt} we show the numerical solution of Eq.~\eqref{EDADME} for two different times when $v_{\alpha}=1/\sqrt{2 \pi}$, $\alpha=1/2$, $a(t)=\exp(Ht)$ with $H=10^{-4}$,  and the initial condition is  $W(x,0)=\delta(x)$.
These solutions (propagators) are qualitatively different from the propagators for a static medium \cite{Metzler2000}.
The main difference is that the maximum of the propagator stays fixed at the origin in the static case whereas it moves when the medium expands. Besides, $W(x,t)$  shows a characteristic cusp at the origin for the static case \cite{Metzler2000}.
However, we see in Fig.~\ref{Fig:DriftWxt} that this effect is much smaller, just a bend at the origin,  when the medium expands.
On the other hand, it should be noted that the propagator is symmetric around the origin where there is no force.  However, we see in  Fig.~\ref{Fig:DriftWxt} that  the propagator has no symmetry when there is an external force.  This means that one cannot obtain this latter propagator from the forceless propagator by means of any kind of displacement. In other words, we see that anomalous diffusion processes under a constant external force in an expanding medium (just as in a static medium \cite{Metzler2000}) are not Galilei invariant.

In Fig.~\ref{Fig:DriftWxt} we have  also included numerical simulation results. In our  simulations all particles start at $x=0$, the jump length pdf $\lambda^*(y)$ is the Gaussian pdf given by  Eq.~\eqref{Gaussianpdf} with $\sigma^2=1/2$, and the waiting time pdf is the Pareto distribution of Eq.~\eqref{varphiSimu}. The external force  we consider induces a probability of jumping to the right $A$ equal to $3/4$, which implies $v_{\alpha}=1/\sqrt{2 \pi}$.  The agreement between simulation results and the numerical solution of Eq.~\eqref{EDADME} is excellent, which provides additional support to the validity of the fractional FPE~\eqref{EDADME}.

\subsubsection{Moments of the propagator for anomalous diffusion under a constant force in an expanding medium}
\label{secMomProADcteF}

Useful information about the diffusive process in an expanding medium can be obtained by evaluating the  first moments of $W(x,t)$. It is possible to find recursive equations for these moments directly from the fractional FPE~\eqref{EDADME}, even when its solution $W(x,t)$ is unknown, by multiplying both members of Eq.~(\ref{EDADME}) by $x^m$ and  integrating the resulting equation over $\mathbb{R}$.
Let us now assume that the force is constant. In this case $v_{\alpha}$ is also constant and from Eq.~(\ref{EDADME}) one finds the recursive relation
\begin{align}
\frac{d}{dt} \langle x^m(t) \rangle =& m (m-1) \frac{\mathfrak{D}_{\alpha}}{a^2(t)} {_0}\mathcal{D}^{1-\alpha}_t \langle x^{m-2}(t) \rangle \nonumber \\
&+ m \frac{v_{\alpha}}{a(t)} ~{_0}\mathcal{D}_t^{1-\alpha} \langle x^{m-1}(t) \rangle.
\label{dxmdt}
\end{align}
For the first moment, one has
\begin{equation}
\label{dxdt}
\frac{d}{dt} \langle x(t) \rangle =  \frac{v_{\alpha}}{a(t)} ~{_0}\mathcal{D}_t^{1-\alpha}\, 1= \frac{v_{\alpha}}{a(t) \Gamma(\alpha)} t^{\alpha-1},
\end{equation}
and hence,
 \begin{equation}
\langle x(t) \rangle = \frac{v_{\alpha}}{\Gamma(\alpha)} \int_0^t \frac{u^{\alpha-1} du}{a(u)}.
\label{xt_v_alpha}
\end{equation}

The equation for the second moment is
\begin{equation}
\frac{d}{dt} \langle x^2(t) \rangle = 2 \frac{\mathfrak{D}_{\alpha}}{\Gamma(\alpha)} \frac{t^{\alpha-1}}{a^2(t)} + 2 \frac{v_{\alpha}}{a(t)} ~{_0}\mathcal{D}_t^{1-\alpha} \langle x(t) \rangle,
\label{dx2t_v_alpha}
\end{equation}
and then
\begin{align}
\langle x^2(t) \rangle =& \langle x^2(t) \rangle_0 + 2 v_{\alpha} \int_0^t \frac{{_0}\mathcal{D}_u^{1-\alpha} \langle x(u) \rangle}{a(u)} du
\label{x2t_v_alpha}
\end{align}
where
\begin{align}
\langle x^2(t) \rangle_0 =2 \frac{\mathfrak{D}_{\alpha}}{\Gamma(\alpha)} \int_0^t \frac{u^{\alpha-1}}{a^2(u)} du
\label{x2t_0}
\end{align}
is the moment of order two when there is no external force.

For  $a(t)=1$ (static medium), Eqs.~\eqref{xt_v_alpha}, \eqref{x2t_v_alpha} and \eqref{x2t_0} become the well-known relations for a static medium \cite{Metzler2000}:
\begin{align}
\langle x(t) \rangle &=\frac{v_\alpha }{\Gamma(1+\alpha)} \, t^\alpha  , \label{xtnoexpa}\\
\langle x^2(t) \rangle &= \langle x^2(t) \rangle_{0} + 2 \frac{[\Gamma(1+\alpha)]^2}{\Gamma(1+2\alpha)} \langle x(t) \rangle^2,
\label{x2minusx2driftless_static} \\
\langle x^2(t) \rangle_0 &= \frac{2\mathfrak{D}_{\alpha}}{\Gamma(1+\alpha)}\, t^\alpha . \label{x2tnoexpa}
\end{align}
From Eqs.~\eqref{xtnoexpa} and \eqref{x2tnoexpa}, and making use of the Stokes-Einstein-Smoluchowski relation~\eqref{SESrelation}, one finds the generalized Einstein relation \cite{Metzler1999PRL,Bouchaud1990,Barkai1998}
\begin{equation}
\label{GEinstein}
\langle x(t) \rangle = \frac{F_0}{2} \frac{\langle x^2(t) \rangle_{0} }{k_B T},
\end{equation}
which relates the first moment in presence of the constant force $F$ to the second moment in absence of this force.
However, for an expanding medium, neither Eq.~\eqref{x2minusx2driftless_static} nor the generalized Einstein relation, \eqref{GEinstein}, holds.

Finally, it should be noted that the variance $2\sigma_x^2(t)= \langle x^2(t) \rangle - \langle x(t) \rangle^2$ when there is an external force, and the variance $\langle x^2(t) \rangle_{0}$ in absence of an external force, are different for an expanding medium as well as for a static medium except if $\alpha=1$ (i.e., except for normal diffusion). This confirms what we saw in   Sec.~\ref{Sec1}, namely, that anomalous CTRWs under an external force field in an expanding medium (as well as in a static medium \cite{Metzler2000}) are not  Galilei invariant.

In the next two sections we obtain explicit expressions for the first two moments for power-law and exponential expansions, and compare them with simulation results.

\subsubsection{Anomalous diffusion, constant force, and power-law expansion}
\label{secAnomaPowerLaw}

Inserting the power-law scaling parameter $a(t)$ given in Eq.~\eqref{aPowLaw} into Eq.~\eqref{xt_v_alpha}, one finds an explicit expression for the first comoving moment:
\begin{equation}
\label{xtpot}
\langle x(t) \rangle =  \frac{v_{\alpha}}{\Gamma(1 + \alpha)} t^{\alpha}\, {_2}F_1 \left( \alpha, \gamma; 1+ \alpha; \frac{-t}{t_0} \right),
\end{equation}
where ${_2}F_1$ is the ordinary hypergeometric function.
From this equation, and taking into account that $y=a(t) x$,  one finds the long-time asymptotic expression of the second moment in physical coordinates \cite{Abramowitz1972}:
\begin{equation}
\langle y (t) \rangle \sim
\begin{cases}
\dfrac{v_{\alpha} }{(\alpha - \gamma) \Gamma (\alpha) }\, t^{\alpha}    &   \text{if}   \quad \alpha > \gamma \\[3mm]
 \dfrac{v_{\alpha}  t^{\gamma} }{\Gamma (\alpha)} \log(t/t_0) &  \text{if} \quad \alpha= \gamma \\[3mm]
\langle x(\infty)\rangle \, (t/t_0)^\gamma &  \text{if}  \quad \alpha < \gamma
\end{cases}
\end{equation}
where $\langle x(\infty)\rangle= v_{\alpha} t_0^{\alpha} \Gamma(\gamma-\alpha)/ \Gamma(\gamma)$.

The second moment $\langle x^2(t) \rangle$ can be evaluated numerically from Eq.~\eqref{x2t_v_alpha} taking into account that
\begin{align}
\frac{{_0}\mathcal{D}_t^{1-\alpha} \langle x \rangle}{\alpha v_\alpha}=&
\frac{2    t^{2\alpha -1} \, _2F_1\left(\alpha ,\gamma ;1+2 \alpha;\frac{-t}{t_0}\right)}{\Gamma(1+2 \alpha)} \nonumber \\
   &-   \frac{   \gamma   t^{2 \alpha } \,{_2}F_1\left(1+\alpha,1+\gamma;2+2 \alpha;\frac{-t}{t_0}\right)}{t_0 \,\Gamma(2+2\alpha)}
\label{x2tPL}
\end{align}
and  \cite{LeVot2017}
\begin{equation}
\label{x2t0PL}
\langle x^2(t) \rangle_0=\frac{2\mathfrak{D}_{\alpha}}{\alpha \Gamma(\alpha)} t^\alpha\,
_2F_1\left(\alpha ,2\gamma ;1+ \alpha
   ;\frac{-t}{t_0}\right).
\end{equation}
Again, from these equations one can get the long-time asymptotic expression of the second moment in physical coordinates  \cite{Abramowitz1972}:
\begin{equation}
\langle y^2 (t) \rangle \sim
\begin{cases}
\dfrac{v^2_{\alpha}    \Gamma(\alpha-\gamma) }{(\alpha - \gamma) \Gamma (\alpha)  \Gamma(2\alpha-\gamma)}  t^{2\alpha} &   \text{if}   \quad \alpha > \gamma , \\[3mm]
 \left[\dfrac{v_{\alpha}  t^{\gamma} }{\Gamma (\alpha)} \log(t/t_0)\right]^2 &  \text{if} \quad \alpha= \gamma , \\[3mm]
\langle x^2 (\infty) \rangle \, (t/t_0)^{2\gamma}&  \text{if}  \quad \alpha < \gamma .
\end{cases}
\end{equation}

These results for the first two moments of the displacement of the particles provides valuable information about the nature of the diffusion-advection process in a power-law expanding medium.
For example, if  $\alpha>\gamma$,  one sees that the first two moments   $\langle y \rangle$ and  $\langle y^2 \rangle$ grow as $t^{\alpha}$ and $t^{2\alpha}$, respectively, for long times,  which is just the way in which these two moments grow in a static medium \cite{Yuste2016}. Thus, we realize that the medium expansion hardly affects the diffusion-advection process in this case. In other words, regarding the spread of the particles, the expansion of the medium is subdominant with respect to the diffusion-advection process if $\alpha>\gamma$. We say that the diffusion of particles  is ``faster'' than the expansion of the medium.
However,  $\langle y \rangle$  and $\langle y^2 \rangle$ grow  as $t^{\gamma}$ and $t^{2\gamma}$ (with  logarithmic corrections in the marginal case $\alpha=\gamma$) if $\alpha<\gamma$,  i.e, the displacement of the particles grows in the same way as the distance between static points does. Thus, we conclude that the  spread of the walkers is mainly driven by the expansion of the medium (i.e., by the Hubble flux) if $\alpha<\gamma$. In this case we say that the expansion of the medium is ``faster'' than the diffusion of particles.

\begin{figure}[t]
\begin{center}
\includegraphics[width=0.45\textwidth]{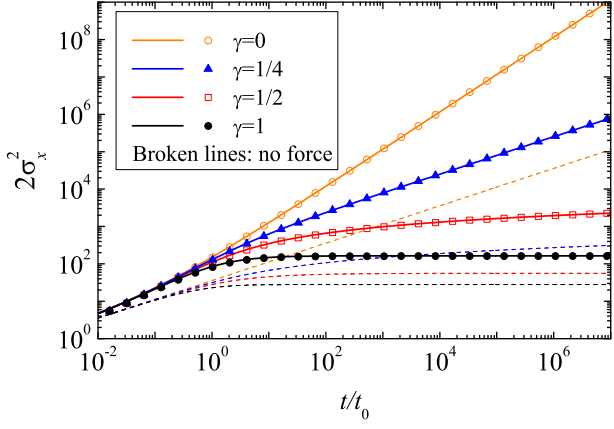}
\end{center}
\caption{Comoving variance for a subdiffusion-advection process ($\alpha=1/2$, and $\mathfrak{D}_{\alpha}=1/2$) in the presence of an external force field ($v_{\alpha}=1/\sqrt{2 \pi}$) in a power-law expanding medium with $t_0=10^3$ and, from top to bottom, $\gamma=0, 1/4, 1/2, 2$. Solid lines represent theoretical values obtained  from Eqs.~\eqref{x2t_v_alpha} and \eqref{x2tPL} whereas broken lines correspond to the force free case ($v_\alpha=0$).  The symbols are simulation results. The random walks were simulated as in Fig.~\ref{Fig:DriftWxt}.}
\label{Fig:VariancePL}
\end{figure}

The  value of  $\langle x^2 (\infty) \rangle$   can be evaluated numerically by means of  Eqs.~\eqref{x2t_v_alpha}, \eqref{x2tPL} and \eqref{x2t0PL}.
In Fig.~\ref{Fig:VariancePL} we compare the variance obtained from Eqs.~\eqref{xtpot},\eqref{x2t_v_alpha},\eqref{x2tPL}, and \eqref{x2t0PL},  with simulation results for $\alpha=1/2$ and four different values of the power-law expansion exponent $\gamma$. The agreement is excellent.
We also show the variance (broken lines) for these same cases when there is no external force.  Notice that for  $t\lesssim t_0$, the expansion of the medium is negligible and the variances hardly depend on the value of the expansion exponent $\gamma$.
At the end of Sec.~\ref{secMomProADcteF} we mentioned that the anomalous diffusion process we are considering, i.e., the CTRW model in an expanding medium,  is not  Galilei invariant. In particular, we noted there that the variance of the propagator  when the particles are subjected to a constant external force is different from the variance when there is no external force. This can be seen in  Fig.~\ref{Fig:VariancePL}: the solid lines (variance for cases with constant external force) and broken lines (variance for cases without external force) are clearly different.

\subsubsection{Anomalous diffusion, constant force, and exponential expansion}

For a medium with exponential expansion, $a(t)=\exp(Ht)$, Eq.~\eqref{x2t_v_alpha} leads to
\begin{align}
\langle x (t) \rangle =& \frac{v_{\alpha}}{ H^{\alpha} } \, \left[1-\frac{\Gamma (\alpha,Ht)}{\Gamma (\alpha )}\right]
\end{align}
where  $\Gamma (\alpha,z)$ is the upper incomplete gamma function.  For long times, one finds that \cite{Abramowitz1972}
\begin{equation}
\label{xt1Exp}
\langle x (t) \rangle\sim v_{\alpha} t_H^{\alpha}\quad \text{if } \;  H>0
\end{equation}
and
\begin{equation}
\label{yt1Exp}
\langle y (t) \rangle\sim -\dfrac{v_{\alpha} t_H }{  \Gamma(\alpha)} t^{\alpha-1}\quad \text{if } \;  H<0.
\end{equation}

The second moment $\langle x (t)^2 \rangle$ can be evaluated  by means of  Eq.~\eqref{x2t_v_alpha} taking into account that
\begin{equation}
 {_0}\mathcal{D}_t^{1-\alpha} \langle x \rangle =
\frac{v_\alpha \sqrt{\pi}}{\Gamma(\alpha)} \left(\frac{H}{t}\right)^{1/2-\alpha} e^{-Ht/2}\, I_{\alpha-1/2}\left(\frac{Ht}{2}\right)
\end{equation}
and \cite{LeVot2017}
\begin{equation}
 \langle x^2(t) \rangle_{0} =2 \mathfrak{D}_\alpha (2H)^{-\alpha} \frac{\Gamma(\alpha,2Ht)}{\Gamma(\alpha)} .
\end{equation}
The function $I_{\nu}(z)$ is the modified Bessel function of the first kind. From these expressions it is possible to find the long-time behavior of the second moment $\langle x^2 \rangle$.
For $H>0$ one has
\begin{align}
\label{x2infHg0}
\langle x^2(\infty) \rangle
 = 2^{1-\alpha} \left(\mathfrak{D}_{\alpha} t_H^{\alpha} + v_{\alpha}^2 t_H^{2\alpha} \right),
\end{align}
whereas
\begin{equation}
\label{yt2Exp}
\langle y^2(t) \rangle \sim \left[-\mathfrak{D}_{\alpha} t_H+\frac{v_\alpha^2 (-t_H)^{1+\alpha}} { \Gamma(\alpha)}\right] t^{\alpha-1}
\end{equation}
for $H<0$.
From these formulas one finds that the variance $2\sigma_x^2(t)$ is different from $\langle x^2(t) \rangle_{0}$ for any $H$, except when $\alpha=1$.  Therefore, the diffusion-advection processes in expanding media described by means the CTRW model, as also happens for static media, are not  Galilei invariant except when the diffusion is normal.

Equations~\eqref{xt1Exp} and \eqref{x2infHg0} tell us that  the two first comoving moments go to a constant value for $t\to\infty$ if $H>0$.  One can use Eq.~\eqref{dxmdt} to prove that this is true for any other moment. Accordingly, the propagator evolves to a well-defined stationary profile in comoving coordinates. Therefore, in the physical space, the particles behave as in the case of a power-law expanding medium where $\alpha<\gamma$ (see Sec.~\ref{secAnomaPowerLaw}), a case where the  displacement of the particles is mainly driven by the expansion of the medium and where the intrinsic movement of the particles (their movement due to their jumps) is negligible.

\begin{figure}[t]
\begin{center}
\includegraphics[width=0.45\textwidth]{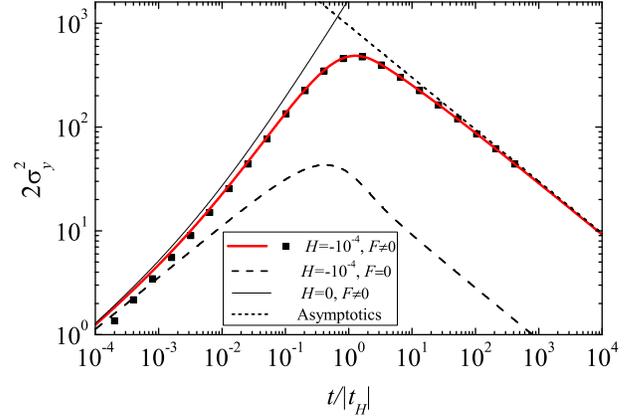}
\end{center}
\caption{Physical variance for a subdiffusion-advection process ($\alpha=1/2$ and  $\mathfrak{D}_{\alpha}=1/2$) under an external force field ($v_{\alpha}=1/\sqrt{2 \pi}$) in an exponential contracting medium with $H=-10^{-4}$.  The symbols are simulation results obtained as described in Fig.~\ref{Fig:DriftWxt}. The thick solid line are theoretical results.  The dashed line corresponds to the case with no external force.   For comparison, we also provide the results for the case with external force but for a static medium (thin solid line).  The short dashed line corresponds to the long-time asymptotic expression obtained from Eqs.~\eqref{yt1Exp} and \eqref{yt2Exp}.}
\label{Fig:VarianceExp}
\end{figure}

\begin{figure}[t]
\begin{center}
\includegraphics[width=0.45\textwidth]{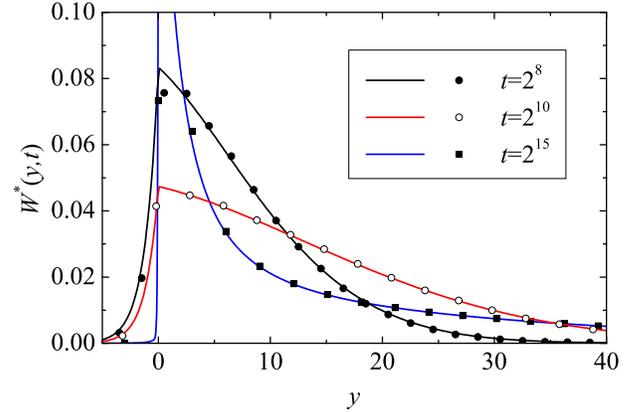}
\end{center}
\caption{Propagator $W^*(y,t)$   for subdiffusive particles ($\alpha=1/2$ and  $\mathfrak{D}_{\alpha}=1/2$) subjected  to an external force field ($v_{\alpha}=1/\sqrt{2 \pi}$) in an exponential contracting medium with $H=-10^{-4}$.   The symbols are simulation results obtained as described in Fig.~\ref{Fig:DriftWxt} for $t=2^8$ (filled circles), $t=2^{10}$ (open circles) and $t=2^{15}$ (squares).  The lines correspond to numerical solutions of Eq.~\eqref{EDADME} by means of the fractional Crank-Nicolson method of Ref.~\cite{Yuste2006} with $\Delta x=\Delta t=0.1$ for $t=2^8$ and $t=2^{10}$, and  $\Delta x=\Delta t=1$ for $t=2^{15}$.}
\label{Fig:PropagatorBigCrunch}
\end{figure}

However, the behavior is completely different when $H<0$.
For example, the first two moments of the physical displacement $y$ go to zero as $t^{\alpha-1}$ for long times [see Eqs.~\eqref{yt1Exp} and \eqref{yt2Exp}].
In fact, it is not difficult to prove by induction that $t^{\alpha-1}$ is the long-time asymptotic behavior of any moment when $H<0$. Let's see it.
From Eq.~\eqref{dxmdt} it is easy to see that the $m$th moment $\langle y^m(t) \rangle = a^m(t) \langle x^m(t) \rangle$ of the physical propagator $W^*(y,t)$  satisfies
\begin{align}
\frac{d}{dt} \langle y^m \rangle =& m (m-1) \mathfrak{D}_{\alpha} a^{m-2} {_0}\mathcal{D}^{1-\alpha}_t \left[ \frac{\langle y^{m-2} \rangle}{a^{m-2}} \right] \nonumber \\
&+ m v_{\alpha} a^{m-1} {_0}\mathcal{D}^{1-\alpha}_t \left[ \frac{\langle y^{m-1} \rangle}{a^{m-1}} \right] + m \frac{\dot{a}}{a} \langle y^m \rangle.
\label{dymdt}
\end{align}
Taking into account that $\langle y^m (0) \rangle = 0$ for the propagator,  Eq.~\eqref{dymdt} is equivalent to
\begin{align}
\label{ymsEq2}
(s-mH) \langle \tilde y^m  \rangle = & \frac{m v_{\alpha}}{ \left[ s - (m-1) H \right]^{\alpha-1}} \langle \tilde y^{m-1}  \rangle  \nonumber \\ &+  \frac{m (m+1) \mathfrak{D}_{\alpha}}{\left[ s - (m-2) H \right]^{\alpha-1}} \langle \tilde y^{m-2} \rangle,
\end{align}
where $\langle \tilde y^m   \rangle\equiv\langle \tilde y^m (s) \rangle$ is the Laplace transform of $\langle y^m (t) \rangle$.
If, for $n=m-1$ and $n=m-2$,  one assumes that   $\langle   y^n (t) \rangle \sim \bar c_{n}t^{\alpha-1}$ when $t\to\infty$,  then $\langle \tilde y^n (s) \rangle \sim  \bar c_n  s^{-\alpha}/\Gamma(\alpha)$ for $s\to 0$.
In this case, from Eq.~\eqref{ymsEq2} one obtains
\begin{equation}
\label{ymsH3}
\langle y^m (s) \rangle \sim \frac{\left[ c_{m,1} (m-1)^{1-\alpha} + c_{m,2} (m-2)^{1-\alpha} \right] }{ m |H|^{\alpha} \Gamma(\alpha)} s^{-\alpha}
\end{equation}
where $ c_{m,1}= mv_\alpha \bar c_{m-1}$ and $ c_{m,2}=\mathfrak{D}_{\alpha} \bar c_{m-2}$.
Equation~\eqref{ymsH3} implies
\begin{equation}
\langle y^m (t) \rangle \sim \frac{ c_{m,1} (m-1)^{1-\alpha} + c_{m,2} (m-2)^{1-\alpha}}{m |H|^{\alpha} } t^{\alpha-1}
\end{equation}
for long times, i.e., $\langle y^m (t) \rangle \to 0$ for $t\to\infty$. This means that the propagator $W^*(y,t)$ goes, eventually, to a Dirac delta function.  In particular, the variance goes to zero for $t\to\infty$.  This is shown in Fig.~\ref{Fig:VarianceExp}. Note that the variance grows initially up to a time around the time $|t_H|$, and then decreases and goes to zero.
This implies that the propagator $W^*(y,t)$ in physical coordinates goes from a peaked  Dirac delta function $\delta(y)$ for $t=0$ to a quite broad function for times around $|t_H|$, and then to an increasingly narrower function that will end up in a Dirac delta function $\delta(y)$ for $t\to\infty$.
In Fig.~\ref{Fig:PropagatorBigCrunch} we can track this behavior: $W^*(y,t)$ is narrower for $t=2^8$ than for $t=2^{10}$, but wider than for $t=2^{15}$ where a noticeable peaked form has already developed.
This behavior is  totally different from  the one found for static media; a case where the particles spread all over the medium. However, it is completely similar  to the one found  for exponentially contracting  media  in the absence of an external force; a behavior called ``big crunch'' in Ref.~\cite{LeVot2017}.

\section{Summary}
\label{Sec_Conclusions}

In this paper we have developed a CTRW model for describing the diffusion properties of normal as well as anomalous diffusive particles that move under an external force field in a uniformly expanding medium. The effect of the force is included by means of a biased jump length distribution in a way similar to the one considered in Ref.~\cite{Metzler1999}.  The expansion of the medium implies the breakdown of the usual formulation of the CTRW model. This is due, essentially, to the fact that the walkers are dragged by the expansion of the medium even when they are resting between jumps. This difficulty is alleviated by the use of~ comoving coordinates instead of the standard physical coordinates.  In this way, we have been able to find a general equation in the Fourier-Laplace space relating the pdf of finding a walker at a given position at a given time to the jump length and waiting time pdfs of the walker. This equation can be written in the form of a generalized FPE if the jump length pdf of the particles has a finite variance. This generalized FPE becomes a fractional FPE when the waiting time pdf is heavy-tailed.

By means of these equations, we have found some interesting results stemming from the interplay between expansion, diffusion and external force.
For normal diffusion, the exact propagator (Green's function) of the generalized FPE can be written in the form of a Gaussian function for any expanding medium. In particular, in a power-law expanding medium with scale factor $a(t)\sim t^\gamma$ for long times, one finds that the spread of particles is dominated by the expansion of the medium when $\gamma>1$, whereas the effect of this expansion is negligible when $\gamma<1/2$.  However, for $1/2<\gamma<1$, the mean position of a particle is determined by the external force whereas its dispersion is determined by the expansion of the medium.  On the other hand, the diffusion process is completely dominated by the expansion of the medium when the  expansion is exponential. Interestingly enough, when the medium contracts exponentially, the propagator reaches a (Gaussian) stationary state with a finite variance.

For anomalous diffusion, it is not easy to find exact solutions of the fractional FPE and we have resorted to a finite-difference fractional Crank-Nicolson method in order to obtain its numerical solutions. In this way, we have found that the propagators in an expanding medium are qualitatively different from the propagators in a static medium.   For example,  the maximum of the propagator is shifted in the course of time for  an exponentially expanding medium, whereas its location stays fixed at the origin for a static medium. These results are supported by simulation results.  We have also provided recurrence equations for the moments of the propagator of the fractional FPE. Thus we have found that the anomalous diffusion process under an external force in an expanding medium is not Galilei invariant and violates the generalized Einstein relation.
For a power-law expanding medium with scale factor $a(t)\sim t^\gamma$ for long times,  we find that the expansion of the medium is not relevant for the spread of the particles if $\alpha>\gamma$, where $\alpha$ is just the anomalous diffusion exponent of the particles.  However, if $\alpha<\gamma$, it turns out that  this spread is mainly  driven  by the expansion of the medium.  For an exponentially expanding medium the behavior of the particles is similar to the latter case, namely, the spread of the particles is essentially accounted for by the expansion of the medium. However, the behavior is completely different for an exponentially contracting medium.  In this case the propagator starts as a Dirac delta function, then becomes a broad function for intermediate times and, eventually, recovers the original form of a Dirac delta function.  This behavior differs from the one we found when the particles are normal diffusive; a case in which  the propagator reaches a stationary state with a finite variance.

The CTRW approach for expanding media can be extended to other problems. For example, we know that, in a static medium, CTRWs with diverging variance lead to L\'evy flights. Thus, a natural generalization is to consider this kind of CTRWs in an expanding medium and to study how the competition between the dilation/contraction of the medium and the divergence of the size of jumps evolves. Another interesting problem would be the obtention of a Galilei invariant diffusion equation for anomalous diffusion in the presence of an external force  \cite{Cairoli2018} when the medium expands.  Finally, it would be interesting to explore the case in which the expansion couples with an external force that depends on the position, e.g., a Hookean force.

\section{Acknowledgments}

This work was partially funded by MINECO (Spain) through Grants No. FIS2016-76359-P (partially financed by FEDER funds) (S.~B.~Y.) and by the Junta de Extremadura through Grant No. GR18079 (S.~B.~Y.). F.~L.~V. acknowledges financial support from the Junta de Extremadura through Grant. No. PD16010 (FSE funds). We also thank Enrique Abad for fruitful discussions and suggestions.

\end{document}